\documentclass[traditabstract]{aa}
\usepackage{epsfig}
\usepackage{natbib}
\usepackage{txfonts}
\bibpunct{(}{)}{;}{a}{}{,} 

\begin{document}

\title{Constraints on the non-thermal emission from $\eta$~Carinae's blast wave of
  1843}
 \author{J.~L. Skilton\inst{1} \and W. Domainko\inst{1} \and J.~A. Hinton\inst{2} \and D.~I. Jones\inst{1} \and S. Ohm\inst{2,3} \and J.~S. Urquhart\inst{4}}

\authorrunning{Skilton et al.}
\titlerunning{Non-thermal emission from $\eta$~Carinae's blast wave}

\institute{Max-Planck-Institut f\"{u}r Kernphysik, PO Box 103980, D 69029 Heidelberg, Germany\\   \email{joanna.skilton@mpi-hd.mpg.de} \and X-ray and Observational Astronomy Group, Department of Physics and Astronomy, University of Leicester, LE1 7RH, UK \and School of Physics and Astronomy, University of Leeds, LS2 9JP, UK \and CSIRO Astronomy and Space Science, P.O. Box 76, Epping, NSW 1710, Australia}

\date{Received; Accepted}

\abstract{Non-thermal hard X-ray and high-energy (HE; 1 MeV $<$ E $<$ 100 GeV)  $\gamma$-ray emission in the direction of $\eta$~Carinae\ has been recently detected using the \emph{INTEGRAL, AGILE} and \emph{Fermi}
 satellites. This emission has been interpreted either in the
 framework of particle acceleration in the colliding wind region
 between the two massive stars or in the very fast moving blast wave
 which originates in the historical 1843 ``Great Eruption''. Archival
 \emph{Chandra} data has been reanalysed to search for signatures of
 particle acceleration in $\eta$~Carinae's blast wave. No shell-like
 structure could be detected in hard X-rays and a limit has been
 placed on the non-thermal X-ray emission from the shell. 
The time dependence of the target radiation field of the Homunculus
is used to develop a single zone model for the blast wave.
Attempting to reconcile the X-ray limit with the HE $\gamma$-ray emission using this model leads to a very hard electron injection spectrum ${\rm d}N/{\rm d}E \propto E^{-\Gamma}$ with $\Gamma <$ 1.8, harder than the canonical value expected from diffusive shock acceleration.}

\maketitle

\section{Introduction}
 
$\eta$~Carinae\ is a binary system composed of a massive primary ($M
\geq90\,M_{\odot}$, $\eta$~Car~A) and a less massive secondary ($M \leq
30\,M_{\odot}$, $\eta$~Car~B) star \citep[see e.g.][]{EtaCar:Nielsen07}
which orbit each other in 2022.7\,$\pm$\,1.3\,days \citep{EtaCar:Damineli08}. $\eta$~Carinae\ experienced a
historical outburst (the ``Great Eruption'') in the 19$^{\rm th}$ century and ejected
$\sim$12\,$M_{\odot}$ of gas which moves outwards at an average speed
of $\sim$650\,km\,s$^{-1}$ \citep{EtaCar:Smith03} forming the
``Homunculus Nebula''. Recent observations show that $\eta$~Carinae\ is
 surrounded at a distance of $\sim$0.25\,pc by a very fast moving
 blast wave ($3500-6000$\,km\,s$^{-1}$) produced in the giant outburst
 of 1843 (also known as the ``Great Eruption'')  \citep{EtaCar:Smith08}.
 This blast wave currently overruns the ``Outer Ejecta'' a ring-like
 structure of material which originates from an ejection of mass from
 $\eta$~Carinae\ $\sim500-1000$ years ago \citep{EtaCar:Walborn78}. This
 fast-moving material mimics a low-energy supernova remnant (SNR)
 shell \citep{EtaCar:Smith08}, with a blast wave moving into the
 ISM with velocities comparable to the historical supernovae RCW\,86
 \citep{Vink:RCW8606} and SN\,1006 \citep{SN1006:Vink05}.

There is evidence for the presence of relativistic particles in
$\eta$~Carinae. The existence of non-thermal X-ray emission was recently
reported by the \emph{INTEGRAL} collaboration
\citep{EtaCar:Integral}. In the high energy (HE; 1\,MeV\,$< E <$\,300\,GeV) domain, a source spatially
coincident with the $\eta$~Carinae\ position was reported by the \emph{AGILE}
and \emph{Fermi} collaborations
\citep{EtaCar:Agile,Fermi:BSL,FERMI:1yr}. Within the measured
positional uncertainties of \emph{INTEGRAL} and \emph{Fermi}-LAT, particle
acceleration via the diffusive shock acceleration (DSA) process is
possible in the colliding wind region (CWR) of $\eta$~Carinae\ and/or in the
expanding blast wave of the Great Eruption of 1843 \citep[see][for a
detailed modelling]{EtaCar:Ohm10}. The absence of significant
variability in the 50\,keV and MeV-GeV regime is surprising in a
colliding wind binary (CWB) picture, particularly during periastron passage where a collapse of the
CWR is expected \citep[see e.g.][and references therein]{Parkin09} and hence no particle acceleration should occur. The
blast-wave scenario provides a good explanation of the observed emission because of the existence of an extended
emission region, thus explaining the lack of significant variability of the source.
While \emph{Fermi} and \emph{INTEGRAL} do not provide
sufficient angular resolution to resolve the blast wave, 
X-ray observations with the current generation of instruments
should provide resolved images of the region. However there is likely
to be some confusion of the possible non-thermal X-ray shell with
thermal emission from the Outer Ejecta \citep[see][for more details]{EtaCar:Seward01}. In this work, archival
\emph{Chandra} X-ray data have been analysed to
search for signatures of accelerated particles in the blast-wave
region.

\section{Chandra data}

\begin{figure}
  \centering
  \begin{tabular}{c}
    \epsfig{file=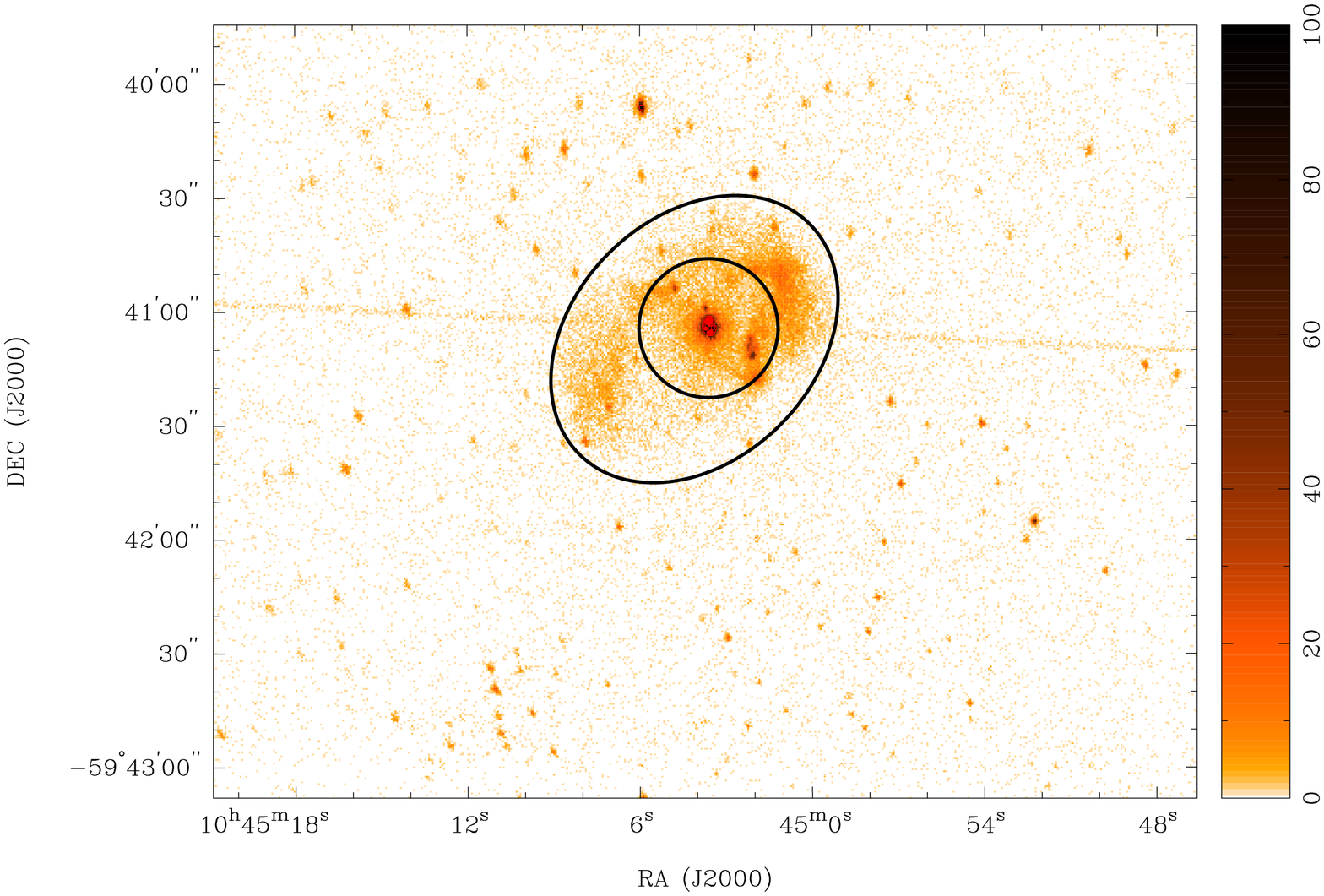, width=1\linewidth, clip=}\\
    \epsfig{file=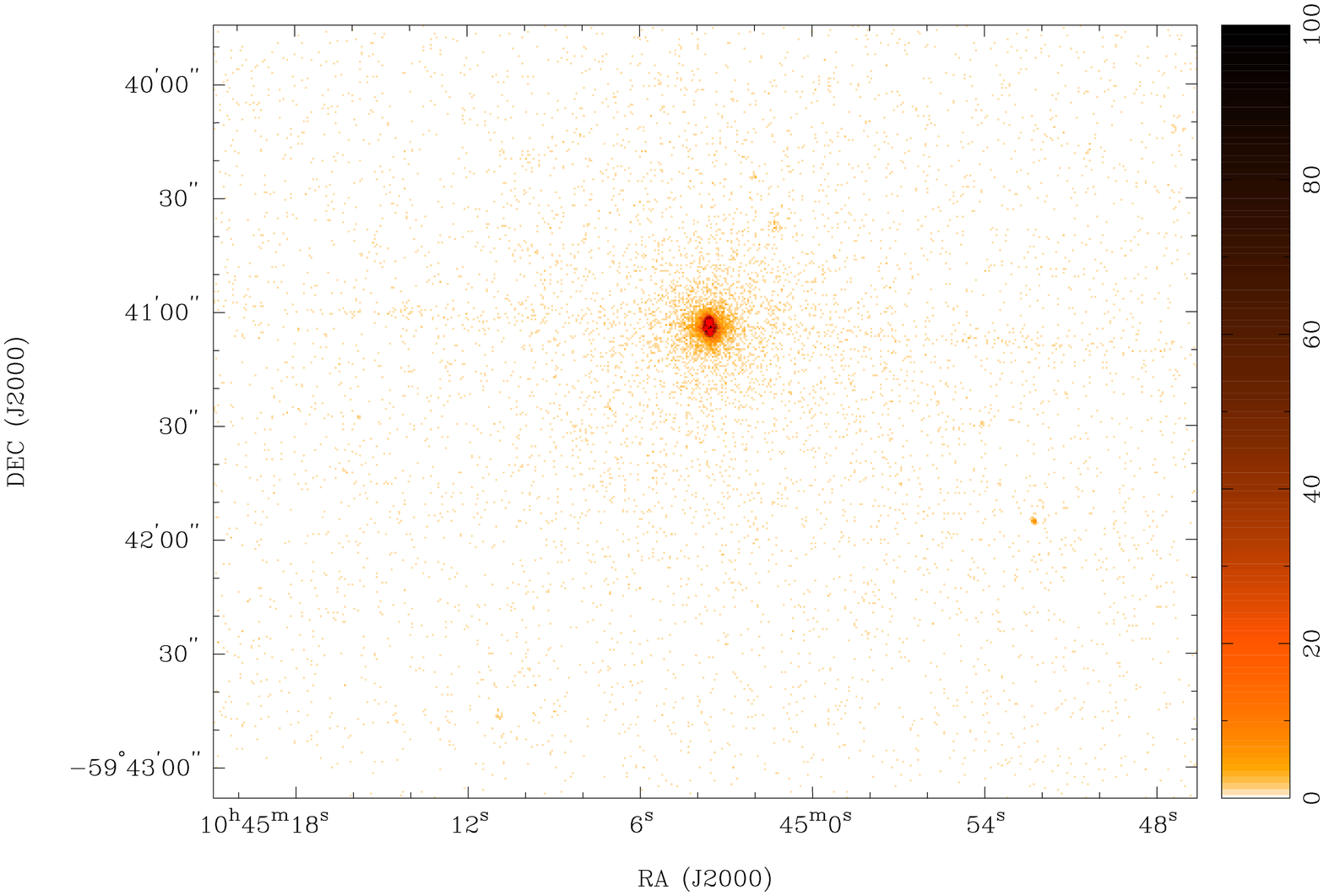, width=1\linewidth, clip=}\\
  \end{tabular}
  \caption{\emph{Chandra} image of the region around $\eta$~Carinae. The top
 (bottom) panel shows the counts map in the energy range 0.3--5.0\,keV
 (5.0--10.0\,keV). The black ellipse and circle show the flux
 extraction region for the ejecta. See text for details.}
  \label{fig:Xrays}
\end{figure}

An 88.2\,ks \emph{Chandra} ACIS-I observation (obs ID 6402) of the Trumpler\,16 region,
taken in August 2006 was reprocessed using the most recent version of
the calibration files. Data processing and reduction was performed
using the \emph{Chandra} CIAO (v4.3) software package and CALDB (v4.4.3). The
data were unaffected by soft-proton flares and so the full observation
time was available for analysis. Standard procedures were followed and
images (count-maps) were created in two energy bands; 0.3\,keV --
5\,keV and 5\,keV -- 10\,keV and are shown in Figure~\ref{fig:Xrays}. The
extended emission attributed to the expanding ejecta around $\eta$~Carinae\
 can be seen clearly in the low-energy image. The high-energy
image however is dominated by the bright point-like emission from
$\eta$~Carinae\  itself. Several other low-significance sources are detected
above 5\,keV, see \citet{EtaCar:Leyder10} for details.

\begin{figure}
  \centering
  \begin{tabular}{c}
    \epsfig{file=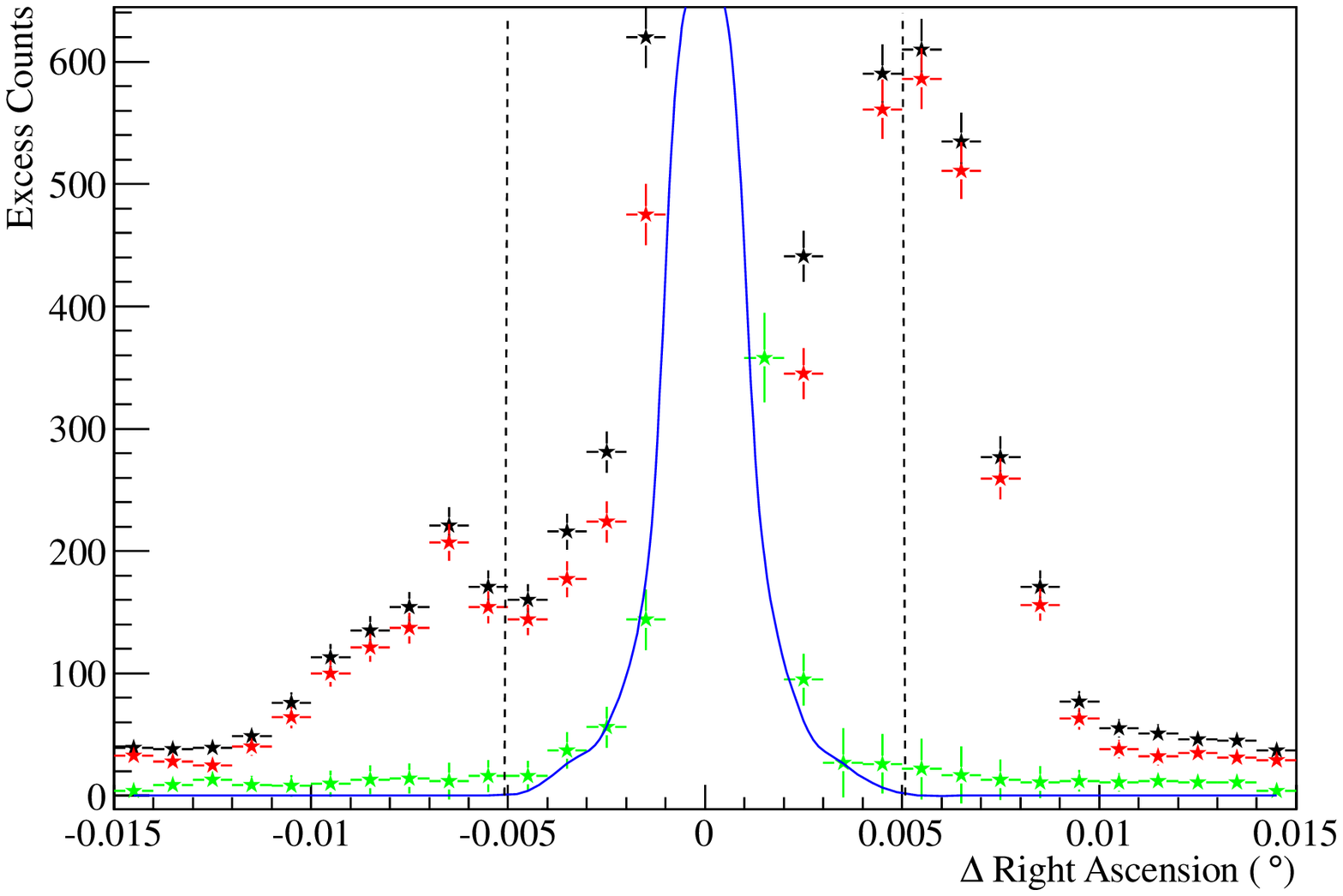, width=1\linewidth, clip=} \\
    \epsfig{file=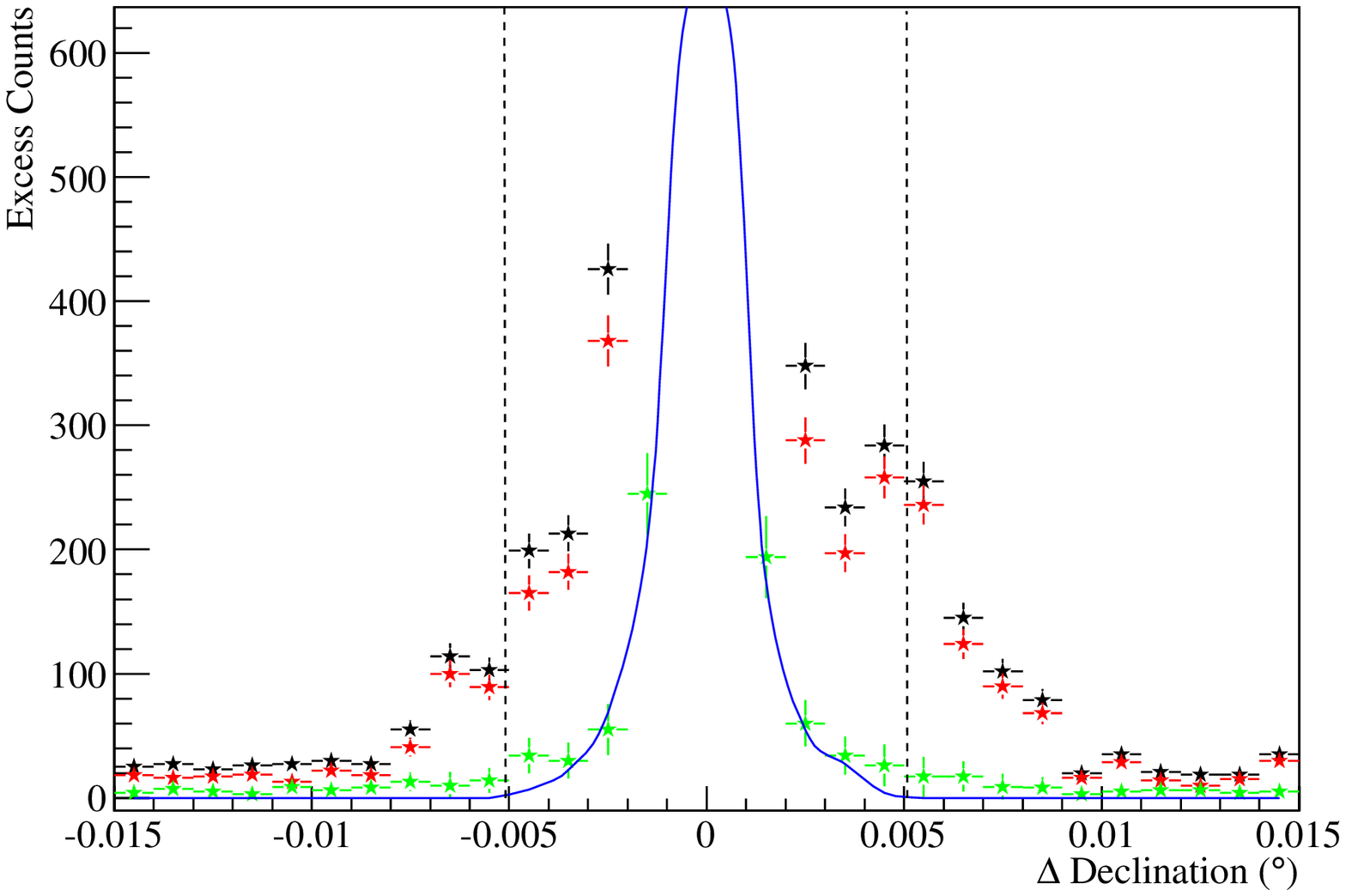, width=1\linewidth, clip=} \\
  \end{tabular}
  \caption{Slices through the \emph{Chandra} images in Right
 Ascension (top panel) and Declination (bottom panel) (see text
 for details). The width of the extraction region was 7$''$. Black
 points represent the profile of the full energy range (0.3--10.0\,keV)
 image. Red points represent the low ($<$5\,keV) energy data and
 green points represent the high ($>$5\,keV) energy data. The blue
 curve shows the simulated \emph{Chandra}-ACIS PSF at the
 chip-position of $\eta$~Carinae\ in this observation. The dashed vertical
 lines show the size of the excluded region around Eta Carinae
 (circular region in Figure 1). The y-scale has been truncated
 to highlight the behaviour away from the central peak. Note that
 the PSF is normalised to the maximum of the full-energy-range data.}
  \label{fig:slices}
\end{figure}

Slices in Right Ascension and Declination were made through the low
and high energy images and through a simulated image of the \emph{Chandra}-ACIS
PSF (created at the same chip-location as $\eta$~Carinae\ in this
observation). The slice regions were 7$''$ wide and the projections
 are shown in Figure~\ref{fig:slices}. It is
clear to see that there is little emission above the PSF in the
high energy map (green points) from the $\eta$~Carinae\ region.

The flux from the shell was estimated from the elliptical region
 shown in Figure~\ref{fig:Xrays}. A circular region (also shown
 in Figure~\ref{fig:Xrays}) centred on the position of $\eta$~Carinae\
 with a radius of 0.305$'$ was excluded from the flux calculation.
 The background was estimated from a large elliptical region
to the south west of the ``on region''. All point sources were
 removed from the background region before extracting the flux.
The background-subtracted flux from the shell in the 5-10\,keV
 band was calculated to be 4.3$\times$10$^{-13}$erg\,cm$^{-2}$s$^{-1}$.
 No spectral analysis of the region has been attempted and so
 this value represents the upper limit on the non-thermal flux
 from the ejecta.

\section{Fermi-LAT data}
 The hypothesis that particle acceleration is occurring in the
 expanding ejecta surrounding $\eta$~Carinae\ suggested by
 \cite{EtaCar:Ohm10} was based on an analysis of the first 11 months
 of \emph{Fermi}-LAT observations. More recent analyses of increased data
 sets are now available and are used here to constrain the model of
 \cite{EtaCar:Ohm10}. 

 An analysis including 21 months of Fermi-LAT data has been presented
 by \cite{EtaCar:Farnier11}. 
 This analysis reveals two components of
 the HE emission; a low-energy part (hereafter L-component) which is
 best described by a power-law with spectral index $\Gamma=1.69\pm
 0.12$ and exponential cut-off at $E_c=1.8\pm0.5$\,GeV, and a second,
 high-energy component (hereafter H-component) extending to
 $\approx100$\,GeV, best described by a pure power law with index
 $1.85\pm0.25$. No temporal variability in either component is
 reported in \cite{EtaCar:Farnier11}. However, some variability
 in the H-component is reported by \cite{EtaCar:Walter11}.
 \citet{EtaCar:Farnier11} have proposed that the Fermi-H
 component results from the interaction of accelerated protons and
 nuclei. This interpretation is attractive in the sense that
 accelerated protons can have a higher maximum acceleration energy
 and suffer less from losses than electrons. We follow this approach
 in this discussion attributing the two components to electrons (L-component)
 and protons (H-component) in the blast wave.

\section{Origin of the HE $\gamma$-ray emission}

The properties of the HE $\gamma$-ray emission from the region in
 and around $\eta$~Carinae\ are extremely challenging to interpret
 within the framework of any current model. In particular,
 for models in which the emission originates in the CWR,
 the lack of observed variability for the bulk of
 the emission is hard to reconcile with the dramatic changes
 seen at other wavelengths during periastron passage \citep[see][and references therein]{Damineli08} and the
 very short cooling time of relativistic particles in the system
 \citep{EtaCar:Farnier11,EtaCar:Bednarek11}. The outer blast-wave
 scenario proposed in \citet{EtaCar:Ohm10} provides a promising
 alternative in the sense that short timescale variability is
 not expected. However, the \emph{Chandra} observations presented
 here place rather tight constraints on this scenario with
 important consequences for shock acceleration in systems of
 this type. Here we present these constraints and the refined
 model of the blast wave emission and discuss the more general
 implications of our results.

\subsection{Improved Blast Wave Model} 

The single-zone, time-dependent numerical model used in
\citet{EtaCar:Ohm10} and described in detail in
\citet{HintonAharonian2007} has been modified here such that the
radiation field energy density of the Homunculus nebula, which had
been assumed to be static is now modelled as a function of time.
The Homunculus is illuminated from the inside by the star $\eta$~Carinae\ with the
(bolometric) luminosity $L_{\eta}$. This starlight is reprocessed by
dust in the optically thick nebula into IR light.
 In the steady-state case, its own luminosity
$L_{\rm H} = L_{\eta}$, but shifted to the IR band. Its thermal time
scale, $t_{\rm th}$, is given by $t_{\rm th} \approx E_{\rm th}/L_{\rm H}$, with
$E_{\rm th}$ being the thermal energy of the Homunculus nebula. For a mass of the
Homunculus nebula of $12\,M_\odot$ and a temperature of 260\,K $E_{\rm th}$ \citep{Gehrz99} is
found to be about $10^{45}$\,erg. Using $L_{\rm H} \approx
10^{40}$\,erg\,s$^{-1}$ \citep{Cox95} leads to a thermal timescale of the
order of a few weeks. This value is much smaller than the age of the
Homunculus nebula and therefore, assuming a steady flux from the star $\eta$~Carinae, a
(pseudo-) steady state is a reasonable assumption. Any luminosity
change of the star $\eta$~Carinae\ would hence be followed by a corresponding
luminosity change of the Homunculus. The temperature of the radiation
field $T$ at the location of the (expanding) ejecta is given by the Stefan Boltzmann law:
$$T = \left(L_{\rm H}/(4 \pi \sigma R^2_{\rm H})\right)^{\frac{1}{4}}\,$$
where, $R_{\rm H}(t)$ is the time-dependent radius of the Homunculus. 
Due to the mass of the expelled material $R_{\rm H}(t)$ follows free
expansion $R_{\rm H}(t) \approx v_{\rm H}\,t\,,$ with $v_{\rm H}$
being the velocity of the ejecta of the Homunculus.
Hence, the time dependent radiation field only depends on the
evolution of the position of the blast wave with respect to $\eta$~Carinae, the luminosity of $\eta$~Carinae\ and
the temperature of the radiation field. Based on the historical light
curve of \citet{EtaCar:Humphreys99}, and the fact that
 the IR emission is simply reprocessed star-light, we assume 
that the integrated luminosity of the nebula in IR is linearly rising
with Eta Car visible magnitude $m_V$.

\subsection{Application to the data}

\begin{figure*}
  \centering
  \epsfig{file=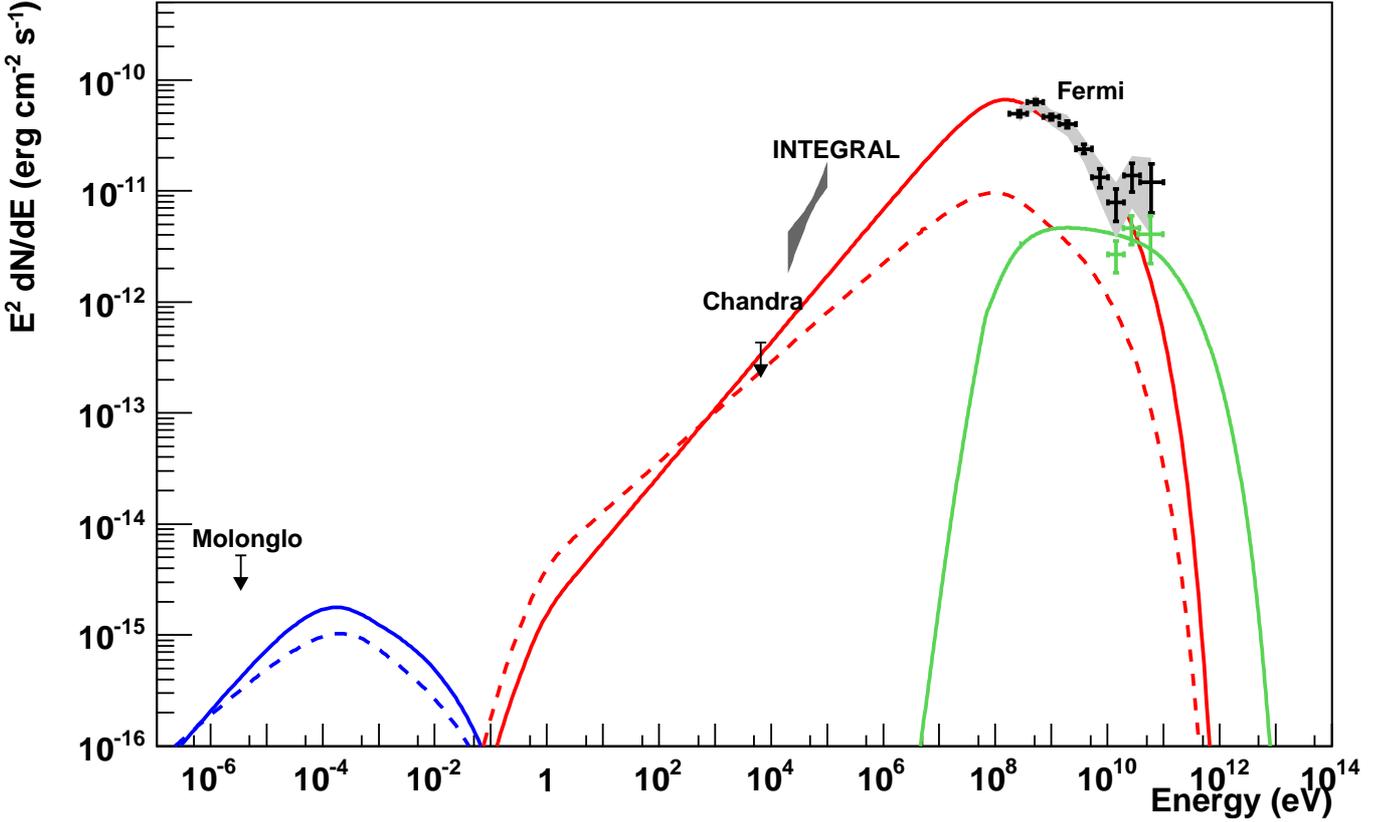, width=1\linewidth, clip=} \\
  \caption{Spectral energy distribution for the region within a few
    arc minutes of $\eta$~Carinae\ as described in \citet{EtaCar:Ohm10}. Curves 
    show a single zone time dependent model for continuous injection of 
    electrons and protons (synchrotron in blue, inverse-Compton
    in red and $\pi^0$-decay in green). A magnetic field strength of $B=10\,\mu$G and 
    electron energy of $E_{\rm e} = 6\times 10^{45}$\,erg is assumed for the 
    model represented by the blue and red solid curves. A spectral index 
    of $\Gamma=1.8$ and maximum electron energy of $E_{\rm max,e} = 
    110$\,GeV is used for this model. $E_{\rm p} = 6\times 10^{48}$\,erg of energy 
    in protons and a maximum energy of $E_{\mathrm{max,p}} = 2$\,TeV is 
    required to reproduce the $\pi^0$-decay $\gamma$-ray component indicated by 
    the green solid curve. Note that the green data points are simply the
    Fermi-LAT data points as presented by \citet{EtaCar:Farnier11}, scaled
    down by a factor of three. The dashed set of curves use a
    model which assume a magnetic field of 
    $B=20\,\mu$G, $E_{\rm e} = 3\times 10^{45}$\,erg of energy in electrons and 
    spectral index $\Gamma=2.1$ with the same maximum electron energy of 
    $E_{\rm max,e} = 110$\,GeV as used before.}
  \label{fig:SED}
\end{figure*}

Fig.~\ref{fig:SED} shows the $\gamma$-ray spectral energy distribution which
is produced by electrons accelerated in the blast wave and
interacting with the time-dependent radiation field of the Homunculus
nebula. The X-ray upper limits presented in this work
restrict the model considerably. 
%
The spectral index of the accelerated electrons is constrained
by the \emph{Chandra} limit to be rather hard ($\Gamma < 1.8$).
This value is much harder than the canonical value ($\Gamma = 2$)
expected from diffusive shock acceleration, but could be realised in very
 strong shocks or in shocks which are modified e.g. by the pressure
 of the accelerated particle population. Using a magnetic field
 strength of 10\,$\mu$G \citep[as used in][]{EtaCar:Ohm10}, the
 total energy in electrons $E_{e}$ for this model would be $6\times
10^{45}$\,erg, representing only a very small fraction ($\approx 10^{-4}$) of
the total kinetic energy $E_{\rm k}$ in the blast wave of $\approx
(4-10)\times10^{49}$\,erg which is in principle available for particle
acceleration.
For a slightly higher magnetic field strength of $20\,\mu$G, and a
spectral index of $\Gamma = 2.1$, the energy in electrons has to be
smaller than $3\times 10^{45}$ in order to agree with the X-ray limit. 
However, this would imply that the vast majority of the
HE $\gamma$-ray emission does not originate in the blast wave. Given these
findings, the fraction of energy in non-thermal electrons compared to
the total kinetic energy in the blast wave is even lower compared to
the model presented before. The association of the soft $\gamma$-ray
emission detected by \emph{INTEGRAL} from the $\eta$~Carinae\ region
\citep{EtaCar:Integral} with the blast wave is problematic due to
the sharp, low energy cutoff required for consistency with the
\emph{Chandra} limits presented both here and in \citet{EtaCar:Leyder10}.
It seems likely that this hard X-ray emission is
associated with the CWB as suggested by \citet{EtaCar:Leyder10} rather
than the blast wave.

\citet{EtaCar:Ohm10} concluded that hadrons were less likely
responsible for the single-component HE $\gamma$-ray emission revealed in
the first 11 months of Fermi-LAT data. This conclusion was derived
from two factors; the presence of HE emission at $\approx 200$\,MeV
(below the $\approx 300$\,MeV threshold energy for $\pi^0$ production)
and the fact that the maximum energy of protons as indicated by the
curvature in the Fermi spectrum lay well below the expected maximum acceleration
energy associated with either the age or size of the system.
It has also been argued \citep{EtaCar:Farnier11,EtaCar:Bednarek11}
that the density of target material for \emph{pp} interaction and
subsequent $\pi^0$-decay $\gamma$-ray production would be too low in the
$\eta$~Carinae\ region.

The new Fermi-LAT data reveal the two-component nature of the HE emission,
and present a good case that the H-component may be of hadronic origin 
\citep{EtaCar:Walter11}. The variability detected in the H-component point to
an origin of at least part of this H-component flux in the colliding wind region.
However it is likely that some part of this H-component originates
from protons accelerated in the blast wave surrounding $\eta$~Carinae. Such emission
would be non-variable and would contribute to the total flux of the
H-component. Scaling the total flux in the H-component down by a factor of three
 \citep[representing the decrease in flux in the high-energy Fermi component
 found by][see green data points in Fig.~\ref{fig:SED}]{EtaCar:Walter11}
 and using the available
target material density of $100$\,cm$^{-3}$ leads to the requirement
of $6\times 10^{48}$\,erg of energy in hadronic cosmic rays in the region.
This represents $6-15$\% of the kinetic energy of the blast wave,
 given the uncertainties in the kinetic energy estimates in the
 blast wave of 4-10 x 10$^{49}$\,erg \citep{EtaCar:Smith03} -- conditions that can reasonably be met.

\section{Summary and conclusions}
We have re-analysed archival \emph{Chandra} data and placed limits on
the non-thermal X-ray emission from the expanding ejecta surrounding
$\eta$~Carinae. The single-zone numerical model of \cite{EtaCar:Ohm10} has
been adapted to account for the time-varying radiation field and to
fit to recent Fermi-LAT HE $\gamma$-ray data. The two-component nature of
the HE emission is best explained by electrons (L-component)
and protons (H-component) respectively. An attempt to
reconcile the new limit on the non-thermal X-ray emission from the shell
with the Fermi L-component data leads to a rather hard electron
injection index of $\Gamma <$1.8. A steady hadronic emission component
originating in the blast wave, at a flux level approximately one third
of the total H-component, can be explained using reasonable numbers for the
energy in cosmic rays, and the target density in the blast wave
region and the maximum proton energy.

\section*{Acknowledgements}

SO acknowledges the support of the Humboldt foundation by a
Feodor-Lynen research fellowship.

\footnotesize{
\bibliographystyle{aa}
\bibliography{EtaCarObs}{}
}
\label{lastpage}

\end{document}